# A Non-orthogonal Distributed Space-Time Coded Protocol
# Part II: Code Construction and DM-G Tradeoff

G.Susinder Rajan and B.Sundar Rajan
Department of Electrical Communication Engineering
Indian Institute of Science, Bangalore, India
Email: {susinder, bsrajan}@ece.iisc.ernet.in

*Abstract*— This is the second part of a two-part series of papers. In this paper, for the generalized non-orthogonal amplify and forward (GNAF) protocol presented in Part-I, a construction of a new family of distributed space-time codes based on Co-ordinate Interleaved Orthogonal Designs (CIOD) which result in reduced Maximum Likelihood (ML) decoding complexity at the destination is proposed. Further, it is established that the recently proposed Toeplitz space-time codes as well as space-time block codes (STBCs) from cyclic division algebras can be used in GNAF protocol. Finally, a lower bound on the optimal Diversity-Multiplexing Gain (DM-G) tradeoff for the GNAF protocol is established and it is shown that this bound approaches the transmit diversity bound asymptotically as the number of relays and the number of channels uses increases.

## I. INTRODUCTION & BACKGROUND

In Part-I [1], a generalized non-orthogonal amplify and forward (GNAF) protocol was proposed and the signal model and code design criteria were obtained. In this paper, we study several code constructions and diversity-multiplexing gain tradeoff for the GNAF protocol. In all the previous works [2]-[7], the important aspect of ML decoding complexity at the destination was not considered. This problem gains significant importance if the *number of relays in the network is large*. An initiative in this direction was first taken in [8], wherein DSTCs with reduced ML decoding complexity were constructed for certain number of relays. Motivated by it, we propose some new code constructions in this paper.

The contributions of this paper are as follows

- Construction of a new family of DSTCs based on co-ordinate interleaved orthogonal designs (CIOD) [10] which have reduced ML decoding complexity compared to the existing ones [5], [8] in the literature. These codes also have low encoding complexity at the relays and are also resistant to relay node failures.

- It is shown that the recently proposed Toeplitz space-time codes [9] can achieve full diversity in the GNAF protocol even with a linear Zero Forcing(ZF)/Minimum Mean Squared Error(MMSE) receiver.

- It is shown that codes from cyclic division algebras [14] can be used as DSTCs in GNAF protocol without any restriction on the rate. Moreover they are also shown to satisfy the requirements for the noise components at the destination to be uncorrelated. Thus the increase in decoding complexity due to implementation of a whitening filter is less, which is not the case with the protocols of [3], [6].

- It is shown that the optimal DM-G tradeoff of the GNAF protocol is better than the protocols of [3], [4] and [5].

The rest of this paper is organized as follows: Several code constructions for the GNAF protocol are presented in Section II and a lower bound on the optimal DM-G tradeoff of GNAF protocol is obtained in Section III. It is assumed that the reader is familiar with the content of Part-I of this paper [1]. The proof for all the theorems and claims are omitted due to lack of space.

## II. CODE CONSTRUCTION FOR THE DISTRIBUTED SPACE-TIME CODED PROTOCOLS

In Part-I [1], it was shown that the simplicity of ML decoding offered by orthogonal designs in colocated MIMO systems was lost when they were used as DSTCs on the relay channel. Towards identifying DSTCs with reduced ML decoding complexity, we define the class of Conjugate Linear Row-Orthogonal DSTCs (CLRO-DSTCs) as those satisfying the following two conditions:

- Condition 1: Any column of the design should contain only the symbols or only the conjugates of the symbols. This in turn implies that each relay is equipped only with a single matrix which we call

- Condition 2: Any relay matrix should satisfy the condition that all its rows are orthogonal to each other.

Note that the class of DSTCs with all the relay matrices being unitary constitute a special case of CLRO-DSTCs.

'relay matrix' as opposed to a pair of matrices as in [1].

## A. Low ML decoding complexity CLRO-DSTCs

Before proceeding to the code construction, we shall briefly overview the notion of $g-$group ML decodable codes [11].
Consider a linear fading MIMO channel

$$Y = SH + W$$

where, $Y \in \mathbb{C}^{T \times N_r}$ is the received matrix, $S \in \mathbb{C}^{T \times N_t}$ is the transmitted codeword, $H \in \mathbb{C}^{N_t \times N_r}$ is the channel matrix and $W \in \mathbb{C}^{T \times N_r}$ is a matrix with all its entries being i.i.d $\mathcal{CN}(0,1)$. Let $N_t$ and $N_r$ denote the number of transmit and receive antennas respectively. Let the transmitted codeword $S$ be obtained from a $T \times N_t$ linear dispersion STBC, $S(X) = \sum_{i=1}^{K} x_i A_i$ in $K$ real variables $x_1, x_2, \ldots, x_K$. The matrices $A_i \in \mathbb{C}^{T \times N_t}$ are called the 'weight matrices' of $S(X)$ and $X = [x_1, x_2, \ldots, x_K]^T \in \mathscr{A} \subset \mathbb{R}^K$ is called the information symbol vector. The maximum likelihood decision rule then minimizes the metric,

$$M(S) \triangleq \min_{S} \| Y - SH \|^2 . \quad (1)$$

Suppose we partition the set of weight matrices into $g$ groups $L_k, k = 1, \ldots, g$, the $k^{th}$ group containing $n_k$ matrices, and also the information symbol vector as $X^T = \left[ X_1^T, X_2^T, \ldots, X_g^T \right]$, where $X_k^T = [x_{j_k+1}, x_{j_k+2}, \ldots, x_{j_k+n_k}]$, $j_1 = 0$ and $j_k = \sum_{i=1}^{k-1} n_i$ for $k = 1, 2, \ldots, g$, then $S(X)$ can be written as

$$S(X) = \sum_{k=1}^{g} S_k(X_k), \text{where } S_k(X_k) = \sum_{i=1}^{n_k} x_{j_k+i} A_{j_k+i}.$$

Now if the information symbols in each group take values independent of information symbols in the other groups and if the weight matrices satisfy

$$A_i^H A_j + A_j^H A_i = \mathbf{0}, \ \forall i \in L_p, \forall j \in L_q, p \neq q. \quad (2)$$

then, it can be easily shown [10], [11] that minimizing the metric in (1) is equivalent to minimizing,

$$M(S)_k = ||Y - S_k(X_k)H||^2 \text{ for each } 1 \leq k \leq g. \quad (3)$$

Such linear dispersion STBCs are said to be $g$-group ML decodable.

The signal model employing a CLRO-DSTC as given in [1] is

$$y = \sqrt{\frac{\pi_3 \pi_1 P^2}{\pi_1 P + 1}} SH + W \text{ where,}$$

$$H^T = \begin{bmatrix} g_0 & \cdots & g_Q f_Q & g_{Q+1} f_{Q+1}^* & \cdots & g_R f_R^* \end{bmatrix}$$

$$W = \begin{bmatrix} w_1 \\ \sqrt{\frac{\pi_3 P}{\pi_1 P+1}} \left( \sum_{i=1}^{Q} g_i A_i v_i + \sum_{i=Q+1}^{R} g_i B_i v_i^* \right) + w_2 \end{bmatrix}$$

$$S = \begin{bmatrix} \sqrt{\frac{\pi_1 P+1}{\pi_3 P}} I_{T_1} s & \cdots & 0 & 0 & \cdots & 0 \\ \sqrt{\frac{\pi_2(\pi_1 P+1)}{\pi_3 \pi_1 P}} A_0 s & \cdots & A_Q s & B_{Q+1} s^* & \cdots & B_R s^* \end{bmatrix} \quad (4)$$

Since the covariance matrix of $W$ is a diagonal matrix and not a scaled identity matrix, we need to apply a whitening filter as shown below.

$$\tilde{y} = \Omega^{-\frac{1}{2}} y = \sqrt{\frac{\pi_3 \pi_1 P^2}{\pi_1 P + 1}} \Omega^{-\frac{1}{2}} SH + \Omega^{-\frac{1}{2}} W$$

Thus the 4 group ML decodability requirement demands that the weight matrices of $\Omega^{-\frac{1}{2}} S$ satisfy (2) for $g = 4$. Let $\Gamma = \left[ \frac{\pi_3 P}{\pi_1 P+1} \left( \sum_{i=1}^{Q} |g_i|^2 A_i A_i^H + \sum_{i=Q+1}^{R} |g_i|^2 B_i B_i^H \right) \right]$. It can be shown that a sufficient condition for $\Omega^{-\frac{1}{2}} S$ to be 4-group ML decodable is that the weight matrices of $\Gamma^{-\frac{1}{2}} \begin{bmatrix} A_1 s & \cdots & A_Q s & B_{Q+1} s^* & \cdots & B_R s^* \end{bmatrix}$ should satisfy (2) for $g = 4$.

In [11], a class of 4-group decodable STBCs have been proposed for the co-located MIMO channel, but these codes are not CLRO-DSTCs and hence are not suitable for GNAF protocols. Therefore there is a need to construct a class of 4-group decodable CLRO-DSTCs. Towards that end, we will first show how Condition 1 of CLRO-DSTC can be partially relaxed. Consider the example of a $4 \times 4$ CIOD [10] which is shown below.

*Example 1:* Let

$$C_1 = \begin{bmatrix} \widetilde{x_1} & -\widetilde{x_2}^* & 0 & 0 \\ \widetilde{x_2} & \widetilde{x_1}^* & 0 & 0 \\ 0 & 0 & \widetilde{x_3} & -\widetilde{x_4}^* \\ 0 & 0 & \widetilde{x_4} & \widetilde{x_3}^* \end{bmatrix} \text{ where,}$$

$$\begin{aligned}
\widetilde{x_1} &= \tfrac{1}{2}\left[(x_1+x_1^*) + (x_3-x_3^*)\right] = x_{1I} + ix_{3Q} \\
\widetilde{x_2} &= \tfrac{1}{2}\left[(x_2+x_2^*) + (x_4-x_4^*)\right] = x_{2I} + ix_{4Q} \\
\widetilde{x_3} &= \tfrac{1}{2}\left[(x_3+x_3^*) + (x_1-x_1^*)\right] = x_{3I} + ix_{1Q} \\
\widetilde{x_4} &= \tfrac{1}{2}\left[(x_4+x_4^*) + (x_2-x_2^*)\right] = x_{4I} + ix_{2Q}.
\end{aligned}$$

The design $C_1$ does not satisfy Condition 1 of CLRO-DSTC with respect to the symbols $x_i, i = 1, \ldots, 4$. But, with respect to the symbols $\widetilde{x}_i, i = 1, \ldots, 4$, the design $C_1$ does satisfy Condition 1 of CLRO-DSTC. Also it can be easily verified that Condition 2 of CLRO-DSTC is also satisfied for the corresponding relay matrices. From Example 1, we see that a STBC can be used in GNAF protocols if by some linear transformation on the variables of the design we obtain a CLRO-DSTC. To be precise, let the CLRO-DSTC be

$$S = \begin{bmatrix} A_1 \tilde{s} & \cdots & A_Q \tilde{s} & B_{Q+1} \tilde{s}^* & \cdots & B_R \tilde{s}^* \end{bmatrix}$$

$$\hat{S} = \begin{bmatrix} A_1 Ps + A_1 Qs^* & \ldots & A_Q Ps + A_Q Qs^* & B_{Q+1} Ps + B_{Q+1} s^* & \ldots & B_R Ps + B_R Qs^* \end{bmatrix} \quad (5)$$

where, $\tilde{s}_{T_1 \times 1}$ is the vector transmitted by the source during the broadcast phase. Suppose $\tilde{s} = Ps + Qs^*$, where $P$ and $Q$ are complex matrices of size $T_1 \times T_1$, then we have $S$ given by (5) shown at the top of this page. Thus we observe that in effect we can use a linear dispersion code that does not satisfy Conditions 1 and 2 of CLRO-DSTC with respect to the variables in the vector $s$. Now we will illustrate how the code given in Example 1 is in fact a 4-group ML decodable code when used in GNAF-II protocol. The corresponding matrix $\Gamma$ for this case is given by

$$\Gamma = \frac{\pi_3 P}{2(\pi_1 P + 1)} \begin{bmatrix} (|g_1|^2 + |g_2|^2) I_2 & 0 \\ 0 & (|g_3|^2 + |g_4|^2) I_2 \end{bmatrix}.$$

It is easy to check that all the weight matrices of

$$\Gamma^{-\frac{1}{2}} \begin{bmatrix} \widetilde{x_1} & -\widetilde{x_2}^* & 0 & 0 \\ \widetilde{x_2} & \widetilde{x_1}^* & 0 & 0 \\ 0 & 0 & \widetilde{x_3} & -\widetilde{x_4}^* \\ 0 & 0 & \widetilde{x_4} & \widetilde{x_3}^* \end{bmatrix}$$ satisfy the required

conditions (2). This is because of the special block diagonal structure of $C_1$ with each block being a replica of the Alamouti code. We shall now generalize the design $C_1$ for any number of relays and call the resultant codes as 'Precoded CIODs' (PCIOD).

**Construction of Precoded CIOD for any number of relays:**

Given an even number $R$, the rate one, $R \times R$ PCIOD $C_P$ is given by (6) shown at the top of the next page. There are totally $2R$ real variables in the design $C_P$. However, we have the following expression from which it is clear that $C_P$ is not fully diverse for arbitrary signal sets.

$$|\Delta C_P^H \Delta C_P| = \left( \sum_{i=0}^{3} \Delta x_i^2 \right) \ldots \left( \sum_{i=2R-4}^{2R-1} \Delta x_i^2 \right)$$

Precoding is to be done in the following manner in order to get full diversity. The $2R$ real variables are first partitioned into 4 groups as follows:

- First group: All real variables whose labeling index is $0 \mod 4$.
- Second group: All real variables whose labeling index is $1 \mod 4$.
- Third group: All real variables whose labeling index is $2 \mod 4$.
- Fourth group: All real variables whose labeling index is $3 \mod 4$.

There will be $\frac{R}{2}$ real variables in each group. Let these $\frac{R}{2}$ real variables in a group take values from a rotated $\mathbb{Z}^{\frac{R}{2}}$ lattice constellation. Moreover, the real variables in a particular group should be allowed to take values independent of the real variables in the other groups. Algebraic number theory provides effective means to construct rotated $\mathbb{Z}^n$ lattices with full diversity and large minimum product distance [12]. Hence by construction, the code $C_P$ is a rate one, full diversity, 4-group ML decodable code. Because of the block diagonal nature of $C_P$ and due to the structure of the corresponding relay matrices, the low ML decoding complexity feature is retained even after applying the whitening filter. If $R$ is odd, then construct a PCIOD for $R+1$ relays and drop the last column to get a $(R+1) \times R$ design.

Note that all the codes constructed in [4], [5], [7] are 1-group ML decodable codes. The codes reported in [8] are 2-group ML decodable CLRO-DSTCs for even number of relays. However, in this paper a class of 4-group ML decodable DSTCs have been provided for any number of relays. We point out two more salient features of PCIODs in the sequel.

*1) Low Encoding complexity at the relays:* Observing the structure of the relay matrices of PCIOD, we note that having received two complex numbers $\begin{bmatrix} a + ib \\ c + id \end{bmatrix}$, a relay should be capable of transmitting either $\begin{bmatrix} a + ib \\ c + id \end{bmatrix}$ or $\begin{bmatrix} -a + ib \\ c - id \end{bmatrix}$, both of which require significantly less complexity as compared to multiplying the received vector by an arbitrary complex matrix.

*2) Resistance to relay node failures:* Note that any two columns of the PCIOD are orthogonal. Hence, even if certain relay nodes fail, which is equivalent to dropping few columns of the designs, the residual diversity benefits are still guaranteed. Thus PCIODs are resistant to relay node failures.

*B. High-rate CLRO-DSTCs*

Till now we have discussed only the case of $T_2 \geq T_1$. The GNAF protocol admits $T_2 \leq T_1$ also by using high rate DSTCs in the cooperation phase. Using high rate DSTCs increases the effective rate as measured by symbols per channel use for every user in the network. For this purpose, we propose DSTCs from cyclic division algebras [14], which are guaranteed to be fully diverse by virtue of the algebraic construction procedure. The design is as shown in (7) at the top of the next page, where $\theta$ is a normalization factor and $f_{i,j}$, $0 \leq i,j \leq (R-1)$ are the variables of the design. It can be easily

$$C_P = \text{diag}\left\{ \begin{bmatrix} x_0 + ix_1 & -x_2 + ix_3 \\ x_2 + ix_3 & x_0 - ix_1 \end{bmatrix}, \ldots, \begin{bmatrix} x_k + ix_{k+1} & -x_{k+2} + ix_{k+3} \\ x_{k+2} + ix_{k+3} & x_k - ix_{k+1} \end{bmatrix}, \ldots, \begin{bmatrix} x_{2R-4} + ix_{2R-3} & -x_{2R-2} + ix_{2R-1} \\ x_{2R-2} + ix_{2R-1} & x_{2R-4} - ix_{2R-3} \end{bmatrix} \right\} \quad (6)$$

$$M_{CDA} = \frac{1}{\sqrt{\theta}} \begin{bmatrix} \sum_{i=0}^{R-1} f_{0,i} t^i & \delta \sum_{i=0}^{R-1} f_{R-1,i} \sigma(t^i) & \delta \sum_{i=0}^{R-1} f_{R-2,i} \sigma^2(t^i) & \cdots & \delta \sum_{i=0}^{R-1} f_{1,i} \sigma^{R-1}(t^i) \\ \sum_{i=0}^{R-1} f_{1,i} t^i & \sum_{i=0}^{R-1} f_{0,i} \sigma(t^i) & \delta \sum_{i=0}^{R-1} f_{R-1,i} \sigma^2(t^i) & \cdots & \delta \sum_{i=0}^{R-1} f_{2,i} \sigma^{R-1}(t^i) \\ \vdots & \vdots & \vdots & \ddots & \vdots \\ \sum_{i=0}^{R-1} f_{R-1,i} t^i & \sum_{i=0}^{R-1} f_{R-2,i} \sigma(t^i) & \sum_{i=0}^{R-1} f_{R-3,i} \sigma^2(t^i) & \cdots & \delta \sum_{i=0}^{R-1} f_{0,i} \sigma^{R-1}(t^i) \end{bmatrix} \quad (7)$$

verified that $M_{CDA}$ is a CLRO-DSTC. In particular, the *standard perfect codes* [15], [16], [17] are also CLRO-DSTCs since they are special cases of those in [14].

### C. CLRO-DSTCs and Non-ML reception

In [9], a class of STBCs called 'Toeplitz STBC' have been shown to achieve full diversity even with a linear Zero Forcing receiver in the colocated MISO (Mulitple input Single Output) channel. A Toeplitz STBC in $T_1$ complex variables $x_1, x_2, \ldots, x_{T_1}$ for $R$ relays is given by

$$\begin{bmatrix} x_1 & 0 & \cdots & 0 \\ x_2 & x_1 & & \vdots \\ \vdots & x_2 & \ddots & 0 \\ x_{T_1} & \vdots & & x_1 \\ 0 & x_{T_1} & \ddots & x_2 \\ \vdots & & \ddots & \vdots \\ 0 & \cdots & 0 & x_{T_1} \end{bmatrix}_{(T_1+R-1) \times R}$$

It is easy to check that Toeplitz STBCs are also CLRO-DSTCs.

*Theorem 1:* The Toeplitz STBC when applied as a DSTC in GNAF-III protocol or Jing-Hassibi protocol [2] employing a square QAM or a PSK signaling scheme of cardinality $K$ and a Linear ZF/MMSE receiver or ML receiver at the destination provides full diversity for the system.

### III. LOWER BOUND ON OPTIMAL DM-G TRADEOFF OF GNAF PROTOCOL

In this section, we establish a lower bound on the optimal DM-G Tradeoff for the GNAF-I,II protocols. Let us first recall certain basic definitions related to DM-G Tradeoff.

*Definition 1 (Multiplexing and diversity gain [13]):* A coding scheme $\{\mathscr{C}(SNR)\}$ is said to achieve *multiplexing gain* $r$ and *diversity gain* $d$ if

$$\lim_{SNR \to \infty} \frac{R(SNR)}{\log SNR} = r \text{ and } \lim_{SNR \to \infty} \frac{\log Pe(SNR)}{\log SNR} = -d$$

where, $R(SNR)$ is the data rate measured by bits per channel use (PCU) and $Pe(SNR)$ is the average error probability using the ML decoder.

Let $d_{opt}(r)$ denote the maximum diversity gain achievable by any coding scheme in GNAF-I protocol for a multiplexing gain of $r$. In Part-I [1], we have shown that all coding schemes for the NAF protocol can be applied in GNAF-I protocol also. Thus the optimal DM-G tradeoff of GNAF-I protocol is lower bounded by the optimal DM-G tradeoff of the NAF protocol. Thus, we have

$$d_{opt}(r) \geq d_{NAF}(r) = R(1-2r)^+ + (1-r)^+.$$

Towards obtaining another lower bound, consider the system model for the GNAF-I protocol when linear STBCs are used.

$$y = \sqrt{\frac{\pi_3 \pi_1 P^2}{\pi_1 P + 1}} SH + W \quad \text{where,} \quad (8)$$

$$S = \begin{bmatrix} \sqrt{\frac{\pi_1 P + 1}{\pi_3 P}} I_{T_1} s & 0 & \cdots & 0 \\ \sqrt{\frac{\pi_2(\pi_1 P + 1)}{\pi_3 \pi_1 P}} A_0 s & A_1 s & \cdots & A_R s \end{bmatrix}$$

$$H^T = \begin{bmatrix} g_0 & g_1 f_1 & \cdots & g_R f_R \end{bmatrix}.$$

Observing (8), we see that it is same as a linear fading channel model with $S$ being the code and $H$ being the channel matrix. The main difference here compared to the colocated MISO channel is that here the channel matrix $H$ has entries which are a product of two independent Gaussian random variables. This channel was named as the 'two product channel' and its optimal DM-G Tradeoff was shown to be the same as the colocated MISO channel in [5], which is

$$d^*(r) = (R+1)(1-r), \quad 0 \leq r \leq 1.$$

Note that the entries of the code $S$ have some constraints to be met which are due to the protocol. For example, certain sub matrices of $S$ have to be zero. The rate of the code $S$ in symbols per channel use is $R_{stc} = \frac{T_1}{T_1+T_2}$. If the submatrix of $S$, given by

$$S_{ER} = \begin{bmatrix} A_1 s & \cdots & A_R s \end{bmatrix}$$

is taken from the perfect code family [15], [16], [17], then it will have the non-vanishing determinant (NVD) property [1]. This in turn implies that the design $S$ will also have the NVD property. It is then immediate from the results of [18], that the DM-G tradeoff of the coding scheme $\{\mathscr{C}(SNR)\}$ obtained using design $S$ and a family of signal sets (usually QAM constellation) when applied in GNAF-I protocol has DM-G tradeoff given by

$$d_{\mathscr{C}}(r) = d^*\left(\frac{r}{R_{stc}}\right) = (R+1)\left(1 - \frac{r(R+1)}{R}\right)^+.$$

A similar result has been obtained for the general colocated MIMO case in the context of study of DM-G tradeoff of Space-Time trellis codes in [19], in which case the rate loss is due to trailing zeros arising due to the need for termination of the trellises. Obviously the DM-G tradeoff of any coding scheme applied in GNAF-I protocol serves as a lower bound for the optimal DM-G tradeoff of the GNAF-I protocol itself. Thus we have

$$d_{opt}(r) \geq (R+1)\left(1 - \frac{r(R+1)}{R}\right)^+.$$

Since the above arguments did not demand any property from the matrix $A_0$, the same lower bound holds good for GNAF-II protocol as well. For values of $r > \frac{R^2}{(R+1)^2 - R}$, we see that the no co-operation case is better in the sense of DM-G tradeoff. Hence for those values of $r$, we can make the source transmit independent symbols in the co-operation phase also. Hence the optimal DM-G tradeoff of GNAF-I protocol is lower bounded as follows:

$$d_{opt}(r) \geq \max\left(1 - r, (R+1)\left(1 - \frac{r(R+1)}{R}\right)^+\right).$$

From the above expression, it can be concluded that the difference between the optimal DM-G tradeoff of GNAF-I protocol and the transmit diversity bound is very negligible for large $R$.

ACKNOWLEDGMENT

This work was partly supported by the DRDO-IISc Program on Advanced Research in Mathematical Engineering, partly by the Council of Scientific & Industrial Research (CSIR), India, through Research Grant (22(0365)/04/EMR-II) and also by Beceem Communications Pvt. Ltd., Bangalore to B.S. Rajan.

---

[1] Determinant of $\Delta S)^H(\Delta S)$ is lower bounded by a constant that is greater than zero and independent of the spectral efficiency.